\documentclass{aa}  
\usepackage{hyperref}
\usepackage{xcolor}
\usepackage{natbib}
\usepackage{aas_macros}

\usepackage{graphicx}
\usepackage{txfonts}
\usepackage{hyperref}
\begin{document}

   \title{Numerical Validation of the Yarkovsky Effect in Super-Fast Rotating Asteroids}
   \author{D. Mar{\v c}eta
          \inst{1}
          \and
          B. Novakovi{\' c}\inst{1} 
          \and 
          M. Gavrilovi{\' c} \inst{2}
          }

   \institute{University of Belgrade, Faculty of Mathematics, Department of Astronomy, Studentski trg 16, 11158 Belgrade, Republic of Serbia \\
         \and
             Astronomical Observatory Belgrade, Volgina 7, 11060 Belgrade, Republic of Serbia\\
             }

   \date{Received July 7, 2025; accepted -}

 
  \abstract
   {Recent discoveries show that asteroids spinning in less than a few minutes undergo sizeable semi-major-axis drifts, possibly driven by the Yarkovsky effect. Analytical formulas can match these drifts only if very low thermal inertia is assumed, implying a dust-fine regolith or a highly porous interior that is difficult to retain under such extreme centrifugal forces.}
   {With analytical theories of the Yarkovsky effect resting on a set of assumptions, their applicability to cases of super-fast rotation should be verified. We aim to evaluate the validity of the analytical models in such scenarios and to determine whether the Yarkovsky effect can explain the observed drift in rapidly rotating asteroids.}
   {We have developed a numerical model of the Yarkovsky effect tailored to super-fast rotators. The code resolves micrometre-scale thermal waves on millisecond time steps, capturing the steep gradients that arise when surface thermal inertia is extremely low. A new 3-D heat-conduction and photon-recoil solver is benchmarked against the THERMOBS thermophysical code and the analytical solution of the Yarkovsky effect, over a range of rotation periods and thermal conductivities.}
   {The analytical Yarkovsky drift agrees well with the numerical solver. For thermal conductivities from $0.0001$ to $1$ $\mathrm{W m^{-1} K^{-1}}$ and spin periods as short as 10 s, the two solutions differ by no more than $15\%$. This confirms that the observed semi-major axis drifts for super-fast rotators can be explained by the Yarkovsky effect and very low thermal inertia. Applied to the 34-s rapid rotator 2016 GE1, the best match of the measured drift is obtained with $\Gamma\lesssim20$ $\mathrm{J m^{-2} K^{-1} s^{-1/2}}$, a value that implies $\sim100$ K temperature swings each spin cycle.}
   {Analytical Yarkovsky expressions remain reliable down to spin periods of a few tens of seconds. The drifts observed in super-fast rotators require low-$\Gamma$ surfaces and might point to rapid thermal fatigue as a regolith-generation mechanism.}


   \maketitle
%

\section{Introduction}

The Yarkovsky effect is a non-gravitational phenomenon produced by the anisotropic emission of thermal radiation that arises from temperature gradients across an asteroid's surface. Solar heating, together with the body's finite thermal inertia, creates those gradients; the delayed re-radiation of energy generates a tiny recoil acceleration. Although this acceleration is far weaker than solar or planetary gravity, it can still drive substantial long-term changes in orbital elements, most notably a secular drift of the semimajor axis \citep{bottke-etal_2006,vok2015}. For these reasons, the Yarkovsky effect has a significant role in the dynamics of small asteroids, and it is an essential component of any accurate prediction of their short- \citep[see, e.g.][]{2021Icar..36914594F} or long-term motion \citep[see, e.g.][]{2022CeMDA.134...34N}.

The magnitude of the effect is a function of the object's size, orbit, and material properties. Though the Yarkovsky effect depends on the asteroid's physical and thermal properties, it manifests itself in the orbital motion. Thus, the mechanism links orbital dynamics, surface composition, and physical parameters in a single framework. Therefore, 
its accurate prediction is essential in various studies of asteroids.

The computation of the Yarkovsky effect typically consists of three stages: (i) determination of surface temperatures via the heat-diffusion equation (HDE), (ii) calculation of the corresponding recoil force, and (iii) the determination of the effect on an asteroid's orbit.

Both analytical and numerical treatments exist. Analytical models deliver closed-form expressions, run almost instantaneously, and the functional dependence on input parameters is transparent. They are, however, first-order approximations.

Depending on how the HDE is solved, thermal models are often grouped into \emph{simple} and \emph{thermophysical} categories \citep{delbo-etal_2015}. Simple models assume the spherical shape and idealized physics (zero conduction, zero roughness) 
\citep[e.g.][]{1989aste.conf..128L,1998Icar..131..291H,2018Icar..303...91M}.  
Thermophysical models incorporate realistic shape and local topography and solve the HDE with finite-conductivity and roughness terms
\citep{1996A&A...310.1011L,2004PhDT.......371D,2007PhDT.......401M,2014Icar..243...58D}. 
Numerical Yarkovsky effect models are typically based on thermophysical models.

Analytical Yarkovsky theories necessarily adopt linearised boundary conditions, spherical shape, and constant thermal parameters, along with circular orbits and uniform rotation \citep{1998A&A...335.1093V,1999A&A...344..362V}. Numerous studies have tested how each simplification influences the result, examining, for example,
(1) non-spherical shape \citep{vokrouhlicky_1998},
(2) variable albedo \citep{2006A&A...459..275V},
(3) the seasonal component for regolith-covered surfaces \citep{vokrouhlicky-broz_1999}, and
(4) nonlinear surface boundary conditions \citep{2013P&SS...84..112S,2014P&SS...97...23S}.  
Those works show that deviations from the zero-order analytical solution are typically of order
20--40~\%, rarely larger, reaching a factor of about two.

The numerical approach allows, in principle, the elimination of all the above-mentioned constraints of the analytical Yarkovsky models. 
Still, they could be very time-consuming, depending on the precision and complexity used in a specific model. Therefore, they are based on the 
thermophysical models of various complexities, and no model so far has accounted for all relevant effects. For instance, most numerical 
models use 1D heat conduction, neglecting lateral conduction.

The numerical Yarkovsky model was first introduced by \citet{2001Icar..149..222S}. Interestingly, it uses 3D heat conduction, 
but was developed for a spherical shape and did not account for surface roughness. The model was used to evaluate the Yarkovsky
effect on the asteroid's orbital semi-major axis, and later on by \citet{2002Icar..156..211S} to evaluate the Yarkovsky effect on orbital
eccentricity and Longitude of perihelion.

Among the first applications of numerical Yarkovsky models were also \citet{chesley-etal_2003}, who developed the fully nonlinear 1D numerical model that incorporates a polyhedral-like shape of the body. The authors applied it for the prediction of the Yarkovsky orbital drift of asteroid (6489) Golevka. \citet{capek2007} has developed a similar numerical model, but extended it to account for variable thermal parameters as well. Both \citet{chesley-etal_2003} and \citet{capek2007} found that for typical asteroids, the analytical solution does not differ much from the numerical results.

\citet{2012MNRAS.423..367R} adapted the so-called Advanced Thermophysical Model (ATPM) described in \citet{2011MNRAS.415.2042R} to simultaneously numerically model the Yarkovsky and YORP effects acting on an asteroid represented by a polyhedral shape model. It includes shadowing, emission vector corrections, thermal-infrared beaming, and global self-heating. The key extension of the ATPM compared to the previous models is that it accounts for thermal-infrared beaming caused by surface roughness. On the other hand, the ATPM is based on 1D heat conduction and does not account for variable thermal parameters. The results obtained by \citet{2012MNRAS.423..367R} show that rough surface thermal-infrared beaming enhances the Yarkovsky orbital drift by typically two tens of percent, although in some specific cases, such as those with very low thermal inertia, it can be as much as a factor of 2.

A growing suite of numerical Yarkovsky models, along with their inherent complexity, has emerged in recent years.  The model of 
\citet{2023Icar..40415647N}, for example, solves full 3-D heat conduction.  Commercial software such as \textit{COMSOL Multiphysics} 
has likewise been adopted: \citet{2022A&A...666A..65X} and \citet{2025arXiv250509482Z} used COMSOL to explore how irregular asteroid 
shapes influence the diurnal component of the effect.  Because such high-fidelity simulations can be slow, \citet{2024A&A...691A.224Z} 
investigated training deep operator neural networks to predict surface temperatures and the resulting Yarkovsky force; the approach 
proved efficient for specific cases (e.g., \ (3200) Phaethon), but its broader applicability remains to be assessed.  Although no single code yet captures every relevant physical process, numerical fidelity continues to improve.  New frameworks are under active 
development, such as the open-source tool proposed by \citet{2024EPSC...17.1121L} with the goal of incorporating additional physics 
while increasing efficiency through optimization and parallelization.

As already outlined above, the especially important role of numerical models is to verify analytical Yarkovsky solutions in 
special cases, which could be most affected by the underlying assumptions of the analytical models. Many such cases have been 
studied, but some remain to be investigated.

For instance, \citet{2001Icar..149..222S} found that for very high eccentricity orbits, where the sign of the Yarkovsky effect 
may be changed, the analytical solution is highly uncertain. This special case may deserve further investigation.

In light of the NASA's DART \citep{2018P&SS..157..104C} and ESA's Hera \citep{2022PSJ.....3..160M} missions, in recent years significant effort has been put in improving the modeling of the Yarkovsky effect for binary asteroids \citep[e.g.][]{2024ApJ...968L...3Z,2024A&A...692L...2Z}, as well as the new thermophysical models specially tailored for this class of asteroids \citep[e.g.][]{2024MNRAS.534.1827J,2025Icar..43416527S}.

Another frontier involves \emph{super-fast rotators}, whose spin periods are measured in seconds to minutes.

\subsection{Extremely fast rotation}

The diurnal Yarkovsky acceleration is governed by the phase lag between maximum insolation and
maximum surface temperature---a lag that first increases with spin rate, boosting the force,
and then decreases once the rotation is so rapid that temperature gradients cannot develop.
The optimum occurs when the dimensionless thermal parameter $\Theta$ is near unity.

Observations have revealed several near-Earth asteroids with rotation periods of 10–200 s that exhibit 
measurable semimajor-axis drifts. If these drifts are Yarkovsky-driven, the inferred surface thermal inertias 
are only a few tens of $\mathrm{J,m^{-2},K^{-1},s^{-1/2}}$ \citep{2021A&A...647A..61F,2023A&A...675A.134F}. 
Such low values are puzzling because neither a fine regolith layer nor a highly porous interior is expected 
to survive the extreme centrifugal stresses of these spin rates.

Analytical Yarkovsky estimates may be unreliable in this regime. Sub-minute rotation induces large temperature 
excursions that challenge the linearised boundary conditions assumed in classical theory \citep{2024PSJ.....5...11N}. 
Consequently, numerical modeling is essential to verify whether the Yarkovsky effect in super-fast rotators can 
indeed produce the observed drifts.

Here, we develop a numerical model that tests whether the Yarkovsky effect alone can reproduce the observed orbital 
drifts of super-fast rotators. The framework solves full three-dimensional heat conduction and an adaptive subsurface grid 
to resolve the shallow thermal skin depth, while assuming a smooth triaxial-ellipsoid shape and neglecting macroscopic surface roughness.

\section{Numerical model}

To calculate the Yarkovsky effect, the thermal state of the asteroid is determined by solving the heat equation using a finite-difference method. The asteroid, modeled as a triaxial ellipsoid, is subdivided into discrete cells, with the assumption that the temperature is uniform within each cell. The rotational state is characterized by the rotation period and the precession period of the rotation axis, which corresponds to one of the main axes of the ellipsoid, enabling the simulation of tumbling motion.

The model is implemented in Python and made available as an open-source package.\footnote{The code is available at \url{https://github.com/dusanmarceta/Yarkovsky}.} It enables the calculation of the isolated diurnal and seasonal components of the Yarkovsky effect. Additionally, it can optionally provide the asteroid’s temperature field, as well as the evolution of the Yarkovsky drift over both the rotation and orbital periods.

These options are provided due to the fact that the diurnal and seasonal components dominate under significantly different physical conditions, most notably depending on the thermal conductivity $k$ and the rotation period of the asteroid. As previously discussed, in cases of very low $k$ and rapid rotation, the penetration depth of the thermal wave becomes extremely shallow, requiring a high vertical resolution and, consequently, a very small time step. Since orbital motion is neglected for the diurnal component, it can be computed at a predefined number of points along the orbit to account for the influence of orbital eccentricity, and the overall effect is obtained as the time-averaged value, computed via numerical integration over the orbital period.

In contrast, the seasonal thermal wave typically penetrates much deeper, which demands discretization of a significantly larger volume of the body, albeit with a much coarser vertical resolution. As a result, the time step can be substantially larger, allowing the seasonal component to be evaluated continuously along the orbit.

\subsection{Numerical grid}

To maintain a smooth and nearly constant cell size across the surface, we utilize a triangular mesh. The grid consists of triangular facets, with the surface resolution defined by the parameter \( a \), which approximately represents the side length of a single facet. The total number of surface cells is approximately $4 S \sqrt{3}/{3 a^2}$, where \( S \) is the surface area of the asteroid. The mesh is structured into parallel, consistently arranged layers, with layer thicknesses progressively increasing from the surface toward the interior of the asteroid.

In order to enable the analysis of the Yarkovsky effect in super-fast rotators, particular attention is given to the vertical grid resolution, as the diurnal thermal wave depth in these cases is very shallow and critically influences the accuracy of the modeling. The thermal wave penetration depth represents the depth at which the temperature variation decreases by a factor of $e$. Diurnal ($l_d$) and seasonal ($l_s$) penetration depths are defined by 

\begin{equation}
l_d = \sqrt{2\pi\frac{ kT_{rot}}{\rho C}}, \quad l_s = \sqrt{2\pi\frac{ kT_{orb}}{\rho C}},
\end{equation}
where $k$ is the thermal conductivity, $C$ is the heat capacity, $\rho$ is the density, and $T_{rot}$ and $T_{orb}$ are rotational and orbital periods, respectively. 

While $\rho$ and $C$ may vary by a factor of a few, thermal conductivity $k$ can differ by as much as 7 orders of magnitude. It is highly dependent on the material properties at the asteroid's surface, ranging from approximately $10^{-5} \, \mathrm{W} \, \mathrm{m}^{-1} \, \mathrm{K}^{-1}$ \citep{2011Icar..214..286K, 2017AIPA....7a5310S} for highly porous regolith to as much as 50 $\mathrm{W} \, \mathrm{m}^{-1} \, \mathrm{K}^{-1}$ \citep{2022M&PS...57.1706N} for materials with high conductivity, such as iron. In addition, Near-Earth asteroids exhibit rotation periods ranging from just a few seconds to several months \citep{2009Icar..202..134W}, leading to significant variations in thermal depth across the population. In the extreme case of a very small $k$ and extremely short rotation period of the order of seconds, the diurnal thermal wave penetration depth ($l_d$) is on the order of a micron. This results in an extremely thin insulating layer on the asteroid’s surface, with a very steep temperature gradient inside it. This requires an extremely high resolution in the vertical direction. In contrast, heat conduction in the lateral (horizontal) direction is much smaller, allowing for a relatively coarser resolution in that direction, which only needs to be fine enough to accurately define the orientation of the surface element with respect to the Sun. 

To ensure full control over the vertical mesh structure, it is defined by three key parameters:
\begin{itemize}
    \item The total thickness of the surface layer being discretized. This is defined by the number of thermal wave penetration depths.
    \item The thickness of the first layer is specified as a fraction of the thermal wave penetration depth.
    \item The total number of layers.
\end{itemize}
The layer thicknesses are then calculated to progressively increase from the surface toward the interior, with the thickness ratio between consecutive layers constant, i.e., ${d_i}/{d_{i-1}} = const$. 

In the adopted grid configuration, each surface cell is surrounded by four neighboring cells, while the interior cells are surrounded by five neighboring cells. The relationships between them, used for simulating heat flow, include the contact surface areas as well as the distances between the centers of mass of these cells. This grid configuration allows adaptive resolution in cases where temperature gradients and consequently heat fluxes vary significantly between the vertical and horizontal directions. Although the mesh is unstructured due to the use of approximately triangular prisms, or more precisely truncated pyramids, it is composed of  consistently arranged parallel layers along the vertical direction, giving the mesh a fully structured character in that axis. The layers share the same connectivity and node arrangement, differing only in their thickness. This structure simplifies the numerical implementation and enhances control over the simulation results.

\subsection{Heat transfer simulation}

To simulate the heat transfer throughout the asteroid, we use the heat transfer equation defining the change in internal energy in a cell, defined as

\begin{equation}\label{eq:heat_equation}
\rho C V \frac{\partial T}{\partial t} = \oint \mathbf{q} \, d \mathbf{s} = \dot{Q},
\end{equation}
where \( \rho \) is the density of the material, \( C \) is the specific heat capacity, \( V \) is the volume of the cell, \( T \) is the temperature, \( \mathbf{q} \) is the heat flux vector, \( d \mathbf{s} \) is the differential area element, and \( \dot{Q} \) represents the total heat flux through the surface.

To solve this equation numerically, we use the Euler method to compute the temperature of the \(i\)-th cell at the \(n+1\) time step as follows:

\begin{equation}
T_i^{(n+1)} = T_i^{(n)} + \frac{\dot{Q}_i^{(n)}}{\rho C \Delta V_i} \Delta t + \mathcal{O}(\Delta t^2),
\end{equation}
where \( \Delta t\) is the time step size, and all physical parameters correspond to the \(i\)-th cell. The critical time step is defined as the time required for heat to propagate the minimum distance among the cells, \( d_{min} \), which is typically the thickness of the surface layer, and is given by:

\begin{equation} 
\Delta t_{cr} = \frac{d_{min}^2 \rho C_p}{k}.
\end{equation}
In order to ensure the stability of the numerical scheme, the time step \( \Delta t \) must be smaller than the critical time step \( \Delta t_{cr} \).

In order to make the model computationally feasible, especially when the thermal wave penetration depth is extremely small, we simulate the heat transfer only in the thin surface layer. The boundary conditions are applied on both sides of this layer, with one side corresponding to the surface of the asteroid and the other side assumed to be at a constant temperature, corresponding to the interior of the asteroid.

The outer boundary condition describes radiative exchange with the environment. Specifically, the heat flux at the surface of the \(i\)-th cell, \( J_i \), is given by:

\begin{equation}
J_i = A_i \left( \max \left( S (1 - \alpha) \, \mathbf{n}_i \cdot \mathbf{r}_{\odot}, \, 0 \right) - \sigma \epsilon T_i^4 \right),
\end{equation}
where \( J_i \) is the external heat flux for surface cell \( i \), \( A_i \) is the area of that cell, \( S\) is the solar heat flux, \( \alpha \) is the surface albedo, \( \mathbf{n}_i \) is the surface normal of cell \( i \), \( \mathbf{r}_{\odot} \) is the unit vector pointing toward the Sun, \( \sigma \) is the Stefan–Boltzmann constant, \( \epsilon \) is the emissivity, and \( T_i \) is the temperature of the surface cell. The expression inside the maximum function represents the absorbed solar flux, which is proportional to the cosine of the angle between the surface normal and the direction of the Sun. When this angle is greater than 90$^\circ$, the surface is turned away from the Sun and should receive no direct insolation. Therefore, the maximum function ensures that only sunlit surfaces receive incoming solar flux, while shaded surfaces receive none.

On the inner side of the layer, we assume that the temperature is constant and equal to the equilibrium temperature, \( T_{eq} \), of a rapidly rotating, isothermal body. This equilibrium temperature is given by:

\begin{equation}\label{eq:equilibrium_temperature}
T_{eq} = \sqrt[4]{\frac{S_{\odot}}{4\sigma r_{hc}^2}},
\end{equation}
where \( S_{\odot}  = 1361 \) W/m$^2$ is the solar constant, and \( r_{hc} \) is the heliocentric distance of the asteroid.

\subsection{Yarkovsky Effect Calculation}

\begin{figure}
\begin{center}
\includegraphics[width=\linewidth]{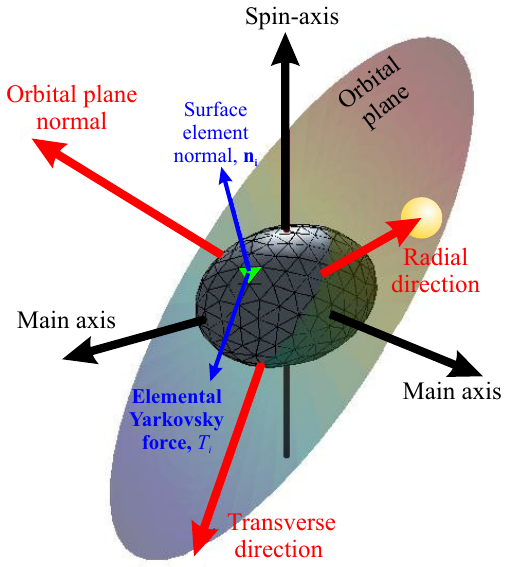}
\caption{Schematic representation of the asteroid model used for the numerical calculation of the Yarkovsky effect, showing the adopted reference frames and the key elements relevant for the computation.}\label{fig:reference frames}
\end{center}
\end{figure}

Since the numerical grid is fixed to the asteroid, all calculations are performed in the asteroid-fixed reference frame, where the apparent motion of the Sun is simulated as a result of the asteroid's rotation, precession of its spin axis, and orbital motion. The resulting radiative forces are computed for each surface element and then summed and projected onto the transverse axis to evaluate the net Yarkovsky effect, as illustrated in Fig. \ref{fig:reference frames}.

The total radiative force acting on each surface element is calculated by summing the elementary forces for each surface cell as follows \citet{2001Icar..149..222S}:

\begin{equation}\label{total_force} 
F = \frac{2 \sigma}{3c} \sum \varepsilon T_i^4 \mathbf{n}_i A_i, 
\end{equation} 
where $c$ is the speed of light, \( \sigma \) is the Stefan–Boltzmann constant, \( \epsilon \) is the emissivity, $T_i$ is the temperature, $\mathbf{n}_i$ represents the unit normal vector of the surface element, and $A_i$ is the outer surface area of the $i$-th surface cell.

The transverse component of the force, representing the projection of \( \mathbf{F} \) onto the transverse direction, is given by $T = \mathbf{F} \cdot \mathbf{t}$ (see Fig.~\ref{fig:reference frames}). Finally, the drift of the semi-major axis is computed from the Gauss planetary equation as \citep[\textit{e.g.}][]{Danby1992}:

\begin{equation}
\frac{da}{dt} = \frac{2a}{n r_{hc}} \cdot \sqrt{1 - e^2} \cdot \frac{T}{m},
\end{equation}
where \( a \) is the semi-major axis, \( n \) is the mean motion, \( r_{hc} \) is the distance from the Sun, \( e \) is the eccentricity, and \( m \) is the mass of the asteroid. To account for the influence of orbital eccentricity on the Yarkovsky drift, the calculation is performed for a defined number of points along the orbit, equally spaced in time, and the drift is obtained as the time-averaged value computed through numerical integration.









\section{Validation of the Numerical Yarkovsky Model}

In this section, we test the validity of our numerical model for determining the Yarkovsky effect.
The validation is performed in two steps. First, in Section~\ref{ss:validation_temperature}, by comparisons of the daily temperature variations in selected regions of a test asteroid, against the results obtained by THERMOBS\footnote{Available at \url{https://www.oca.
eu/images/LAGRANGE/pages_perso/delbo/thermops.tar.gz}}, the widely used thermophysical model by \citet{2004PhDT.......371D}. Once the temperature variation pattern is verified, we switch to comparison of the Yarkovsky effect induced drift in the semi-major axis against an analytical solution (Section~\ref{ss:yarko_validation}).

\subsection{Validation of daily temperature variations against a thermophysical model}%
\label{ss:validation_temperature}

As the first step in our validation procedure, we compared the diurnal surface temperatures returned by our \emph{3-D HDE solver} with those obtained from the well-established thermophysical model THERMOBS. Both codes were run under an identical physical and orbital setup summarised in Table~\ref{tab:params_validation}. Except stated otherwise, both thermal solvers employ a vertical grid of
\(N=64\) layers, each $0.1 l_d$ thick, where $l_d$ is the diurnal
thermal skin depth. Two representative facets are examined: one at the equator, receiving maximum insolation, and another at a latitude of \(45^{\circ}\), chosen to represent intermediate illumination conditions.

\begin{table}[h]
  \centering
  \caption{Input parameters common to both thermal models.}
  \label{tab:params_validation}
  \begin{tabular}{lcl}
    \hline
    Parameter                       & Symbol & Value \\ \hline
    Diameter                        & \(D\)  & \(100\ \mathrm{m}\) \\
    Rotation period                 & \(P\)  & \(5\ \mathrm{hr}\) \\
    Bulk density                    & \(\rho\) & \(2000\ \mathrm{kg\,m^{-3}}\) \\
    Specific heat capacity          & \(C_{p}\) & \(800\ \mathrm{J\,kg^{-1}\,K^{-1}}\) \\
    Bond albedo                     & \(A\) & \(0.10\) \\
    Emissivity                      & \(\epsilon\) & \(0.95\) \\
    Semi–major axis                 & \(a\) & \(1\ \mathrm{au}\) \\
    Orbital eccentricity            & \(e\) & \(0\) \\
    Obliquity                       & \(\gamma\) & \(0^{\circ}\) \\ \hline
  \end{tabular}
\end{table}

\begin{figure}
\begin{center}
\includegraphics[width=\linewidth]{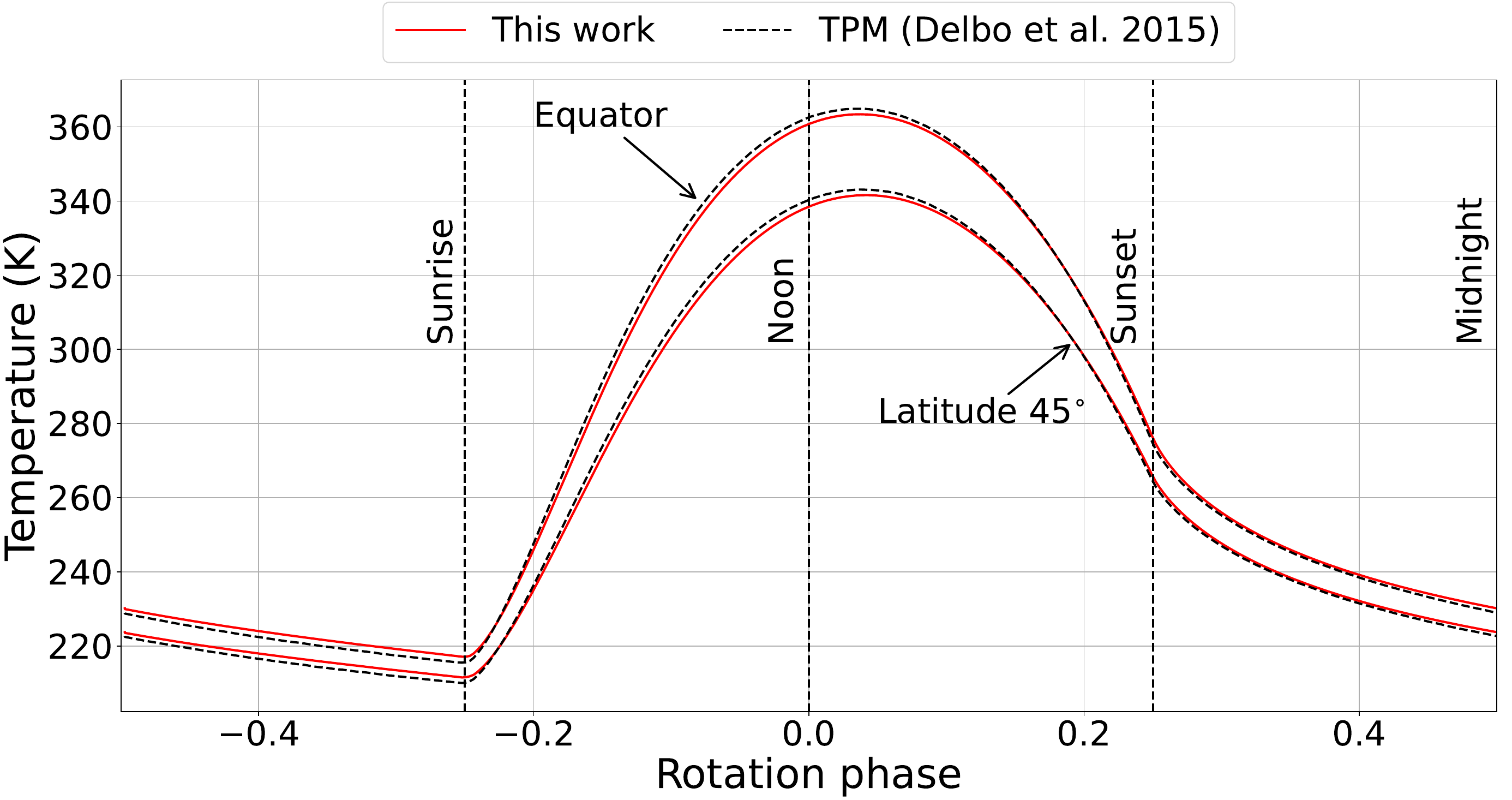}
\caption{Comparison between TPM \citep{delbo-etal_2015} and our model applied to a spherical test asteroid with parameters: $D = 100$ m, $P = 5$ h, $\rho = 2000$ kg/m$^3$, $C_p = 800$ J/kg K, albedo = 0.1, $\epsilon = 0.95$, $a = 1$ au, $e = 0$, and $\gamma = 0^\circ$}\label{fig:validation delbo}
\end{center}
\end{figure}

The obtained results are shown in Figure~\ref{fig:validation delbo}.
Across the full rotation cycle, the two temperature curves are visually almost indistinguishable for both latitudes; the noon peaks, nighttime minima, and the shoulder regions at sunrise and sunset all nearly overlap. A closer inspection reveals that our model predicts marginally warmer temperatures just before sunrise and after sunset, whereas THERMOBS gives slightly hotter temperatures near local noon. In both cases, the mismatch remains below 3~K during daylight and below 2~K at night, which is well under $1\%$ of the diurnal temperature range.

Additional quantitative metrics can be summarized in the following: (i) Root-mean-square (RMS) difference:  $<1\ \mathrm{K}$ at the equator and $<1.5\ \mathrm{K}$ at \(45^{\circ}\), corresponding to \(<0.7\%\) of the \(220\text{–}360\ \mathrm{K}\) dynamic range; (ii) Peak-to-peak amplitude ratio: \(0.998\) (equator) and \(0.996\) (mid-latitude);
(iii) Phase lag of the afternoon maximum: \(<0.01\) rotation cycles (\(<3\ \mathrm{min}\)).

With a single, fully specified parameter set, our solver reproduces the one-dimensional Delbo TPM to better than \(1\%\) in amplitude and a few minutes in phase at both equatorial and mid-latitude locations. This validates the local-scale thermal component of our Yarkovsky
framework and provides a solid foundation for the subsequent recoil force level verification presented in Section~\ref{ss:yarko_validation}.


\subsection{Validation against the analytical Yarkovsky model}
\label{ss:yarko_validation}

We next validate the recoil–force module by comparing the semimajor-axis drift $da/dt$ predicted by our model with the
closed-form diurnal Yarkovsky expression developed by \citet{1998A&A...335.1093V,1999A&A...344..362V}. 

We first investigated how the analytical and numerical results depend on thermal conductivity. Using the same spherical test body as in the temperature check, we sweep the thermal conductivity over $k = 10^{-5}$–$100\, \mathrm{W\,m^{-1}K^{-1}}$ and compute $da/dt$ for three vertical grids that differ only in the number of layers and their thickness expressed via thickness of the top layer ($\Delta_{1}=0.25$, 0.10, and 0.05~$l_d$; 16, 24, and 32 layers in total). 

\begin{figure}
\begin{center}
\includegraphics[width=\linewidth]{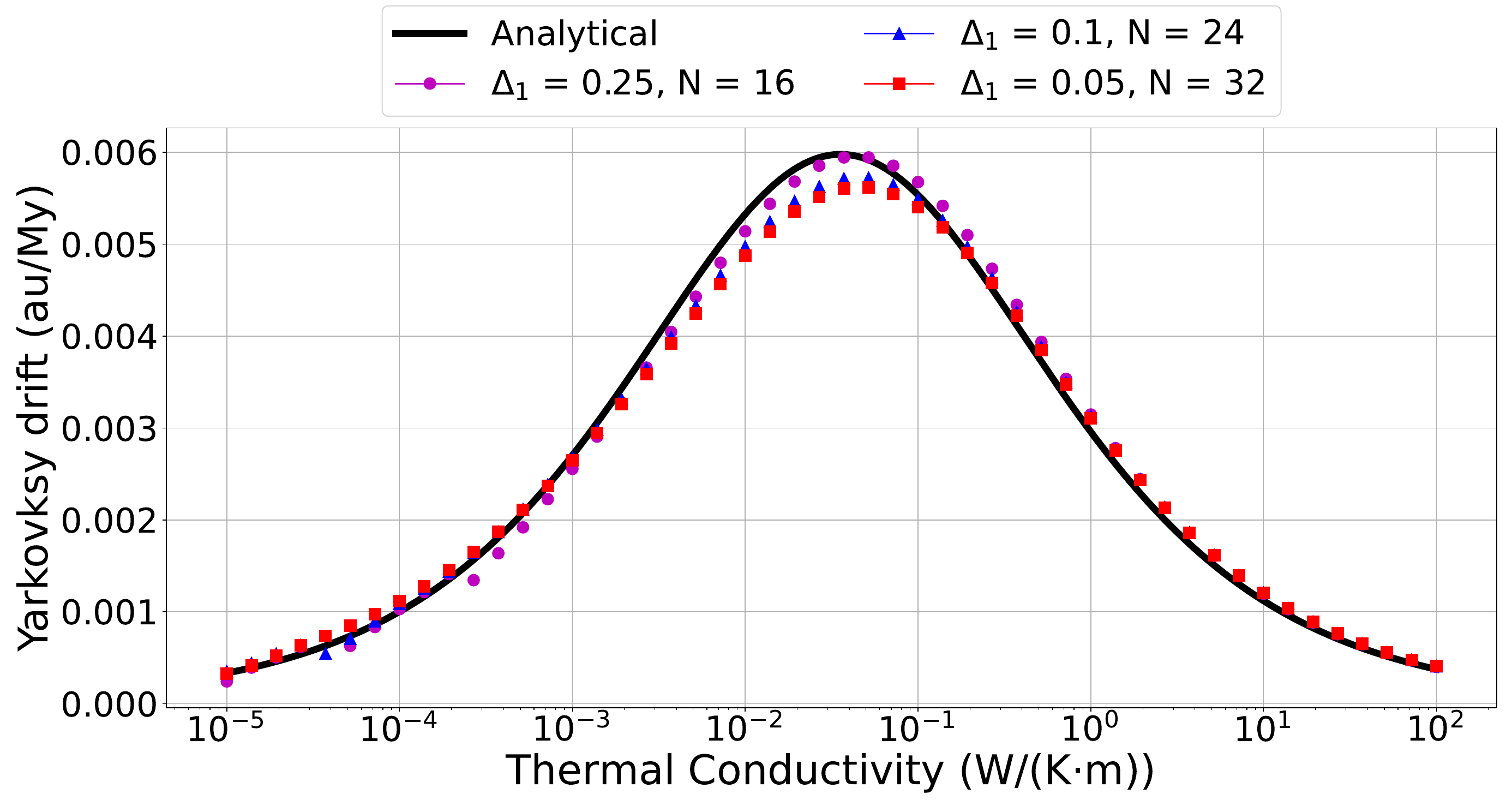}
\caption{Comparison between the analytical and numerical solutions for varying vertical grid resolutions (as indicated in the plot), applied to a test asteroid with parameters listed in Table~\ref{tab:params_validation}.}\label{fig:validation k}
\end{center}
\end{figure}

The resulting numerical curves are plotted alongside the analytical solution in Figure~\ref{fig:validation k}.
The overall agreement between the results obtained with the analytical and numerical models is very good, with a maximum difference of less than $10\%$. The results from the numerical models indicate that vertical grid resolution is relevant; however, the drift converges rapidly with refinement. Over the considered range of conductivities, a grid with $\Delta_1=0.10 l_d$ and N=24 layers matches the finer layer grid ($\Delta_1=0.05 l_d$, N=32) within $<2\%$.

\begin{figure}
\begin{center}
\includegraphics[width=\linewidth]{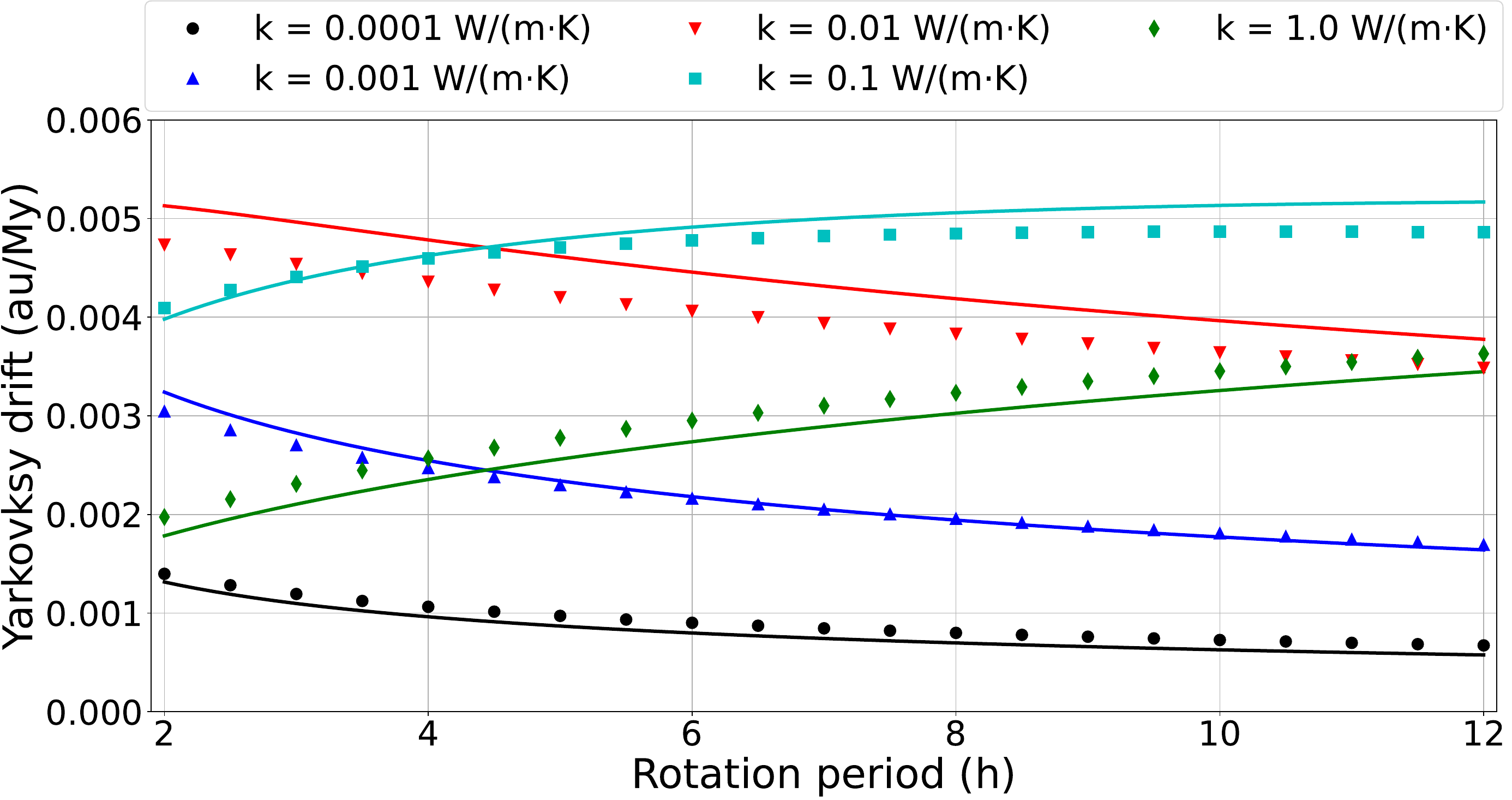}
\caption{Comparison between analytical solutions (solid lines) and numerical results (dots) for different rotation periods, using a grid resolution of 32 layers and a surface layer thickness of $0.05\cdot l_d$. The model is applied to the test asteroid with parameters given in Table~\ref{tab:params_validation}, but using $\gamma = 30^\circ$}\label{fig:validation P}
\end{center}
\end{figure}

Next, we compare the results from the analytical and numerical diurnal Yarkovsky models for rotation periods ranging from 2 to 12 hours, and for five discrete values of thermal conductivity, as shown in Figure~\ref{fig:validation P}. Again, the agreement between the two models is excellent. The nominal deviation and its sign depend on the thermal conductivity, as expected. However, the two models always agree within $10\%$. This is also in agreement with the comparison of numerical and analytical Yarkovsky models presented by other authors \citep[e.g.,][]{chesley-etal_2003,capek2007}. Indeed, \citet{capek2007} found potentially larger deviations between the models for thermal conductivity below $\sim10^{-5}$ $\mathrm{W\,m^{-1}K^{-1}}$, but this is outside the range we studied here.

\begin{figure}
\begin{center}
\includegraphics[width=\linewidth]{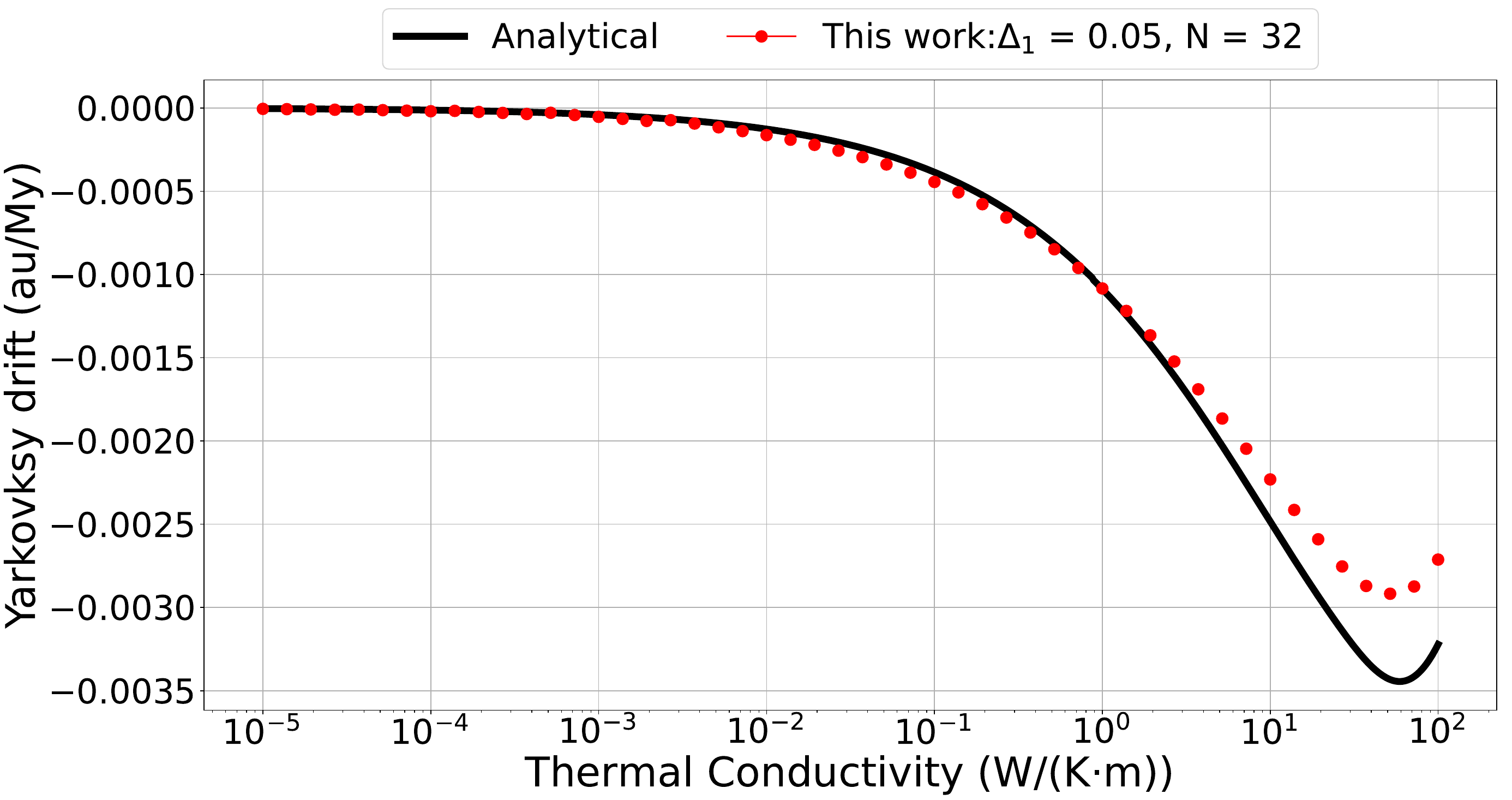}
\caption{The seasonal Yarkovsky drift for the test asteroid determined from the analytical (black line) and numerical (red dots) model.}\label{fig:validation_k_seasonal}
\end{center}
\end{figure}

Finally, we compare the numerical and analytical models for the seasonal component of the Yarkovsky effect. The comparison is performed for the same test asteroid, and by varying thermal conductivity in the $k = 10^{-5}$–$100\, \mathrm{W\,m^{-1}K^{-1}}$ range. The results are shown in Figure~\ref{fig:validation_k_seasonal}. We observe excellent agreement between the two models for $ k \leq 10$. Deviations emerge for $k>10$ $\mathrm{W\,m^{-1}K^{-1}}$, probably driven by analytical simplifications becoming less accurate at high conductivity. Nonetheless, the discrepancy remains below $20\%$ for all tested conductivities.

\section{Results and Discussion}


Having successfully validated our numerical model against both diurnal surface temperature variations and the resulting Yarkovsky drift under typical spin conditions, we now reverse our approach. Given that our numerical model accurately reproduces the analytical results for standard rotation periods and does not share the simplifying assumptions of the analytical formulation, it is expected to maintain its accuracy even under extreme conditions. Therefore, we utilize our numerical model to evaluate the accuracy and limitations of the analytical Yarkovsky model when applied to rapidly rotating asteroids.

\begin{figure}
\begin{center}
\includegraphics[width=\linewidth]{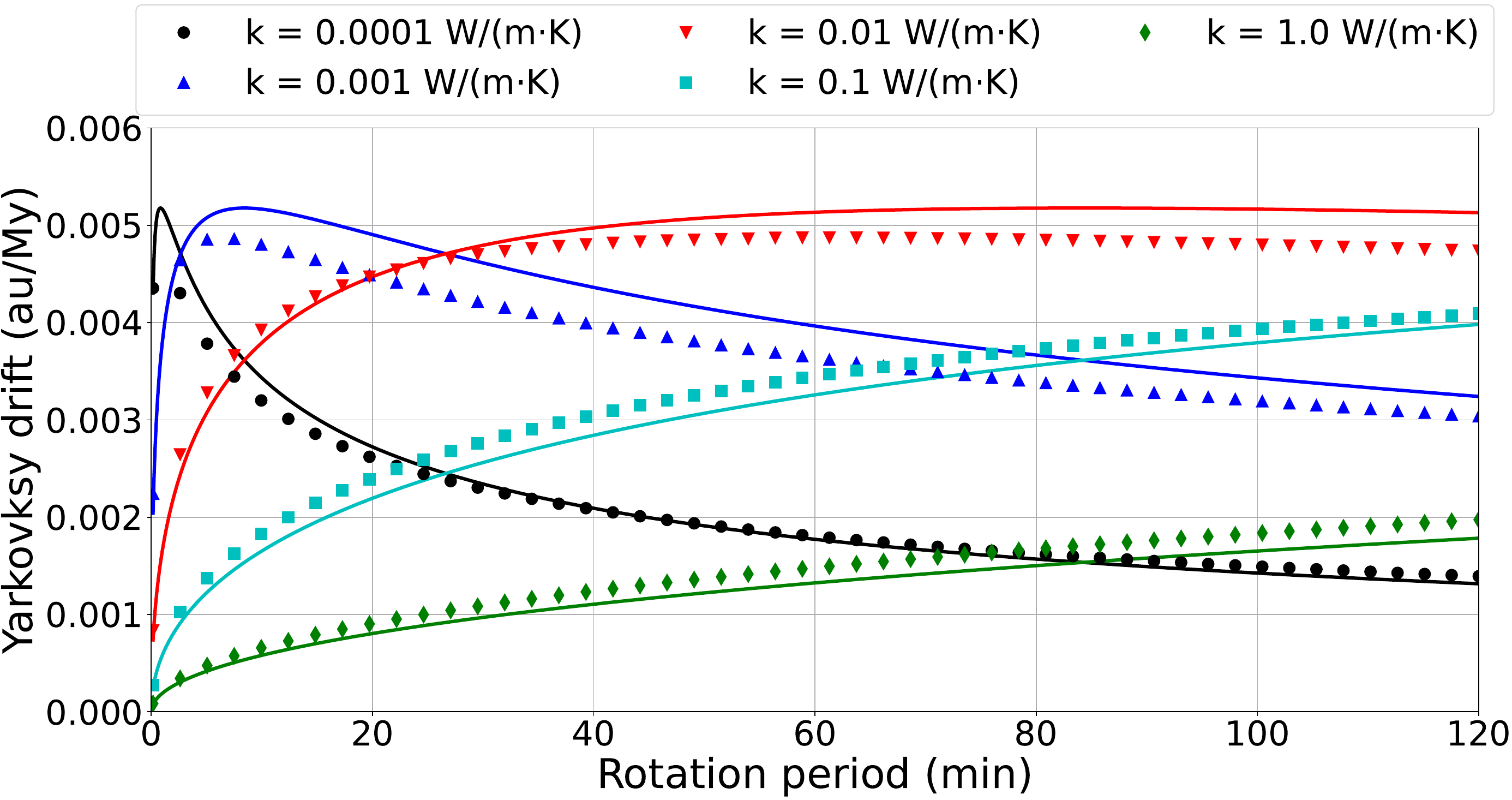}
\caption{Comparison between the analytical and numerical solutions for fast rotation applied to a fictitious asteroid with parameters given in Table \ref{tab:params_validation}.}\label{fig:comparison fast rotation}
\end{center}
\end{figure}

Figure~\ref{fig:comparison fast rotation} presents the comparison between numerical and analytical predictions for rotation periods ranging from 10~s to 2~hr, covering thermal conductivity values from \(10^{-4}\) to \(1\,\mathrm{W\,m^{-1}K^{-1}}\). The overall agreement remains very good, with relative differences ranging from approximately \(-9\%\) to \(+13\%\). These results demonstrate that the analytical Yarkovsky model, despite its inherent simplifying assumptions, remains robust and applicable even under these extreme rotation conditions.

On this basis, we can state that, even at extremely high rotation rates, the observed semi-major axis drifts are achievable through the Yarkovsky effect. The question is which component, diurnal or seasonal, provides the dominant contribution. The diurnal term, generally the stronger of the two, is the obvious candidate. Yet, under rapid rotation, it matches the measured drifts only if the surface thermal inertia is very low. Alternatively, the seasonal component may account for the observed Yarkovsky drift, eliminating the need to invoke special surface features, such as insulating regolith whose survival under extreme centrifugal forces is doubtful. To test this possibility, we mapped both diurnal and seasonal Yarkovsky contributions over the entire physically reasonable parameter space (Table~\ref{tab:PT_GE1}) for two representative fast rotators, 2011 PT and 2016 GE1.

\begin{table*}[]
    \caption{The maximum plausible range for the parameters of asteroids 2011 PT and 2016 GE1 relevant for the magnitude of the Yarkovsky effect.}
    \centering
    \begin{tabular}{l|cc|cc}
        & \multicolumn{2}{c|}{2011 PT} & \multicolumn{2}{c}{2016 GE1} \\
        \hline
        Parameter & Min. value & Max. value & Min. value & Max. value \\
        \hline
        Rotation period (s) & 612 & 612 & 17 & 340 \\
        Diameter (m) & 20 & 120 & 6 & 41 \\
        Density (kg/m$^3$) & 1000 & 3500 & 1000 & 3500 \\
        Bond albedo  & 0 & 0.4 & 0.016 & 0.173 \\
        Thermal conductivity (W/m$\cdot$K) & 10$^{-5}$ & 100 & 10$^{-5}$ & 100 \\
        Heat capacity (J/kg$\cdot$K) & 600 & 1200 & 600 & 1200 \\
        Spin-axis obliquity (deg) & 90 & 180 & 90 & 180\\
        \hline
    \end{tabular}
    \label{tab:PT_GE1}
\end{table*}

\begin{figure}
\begin{center}
\includegraphics[width=\linewidth]{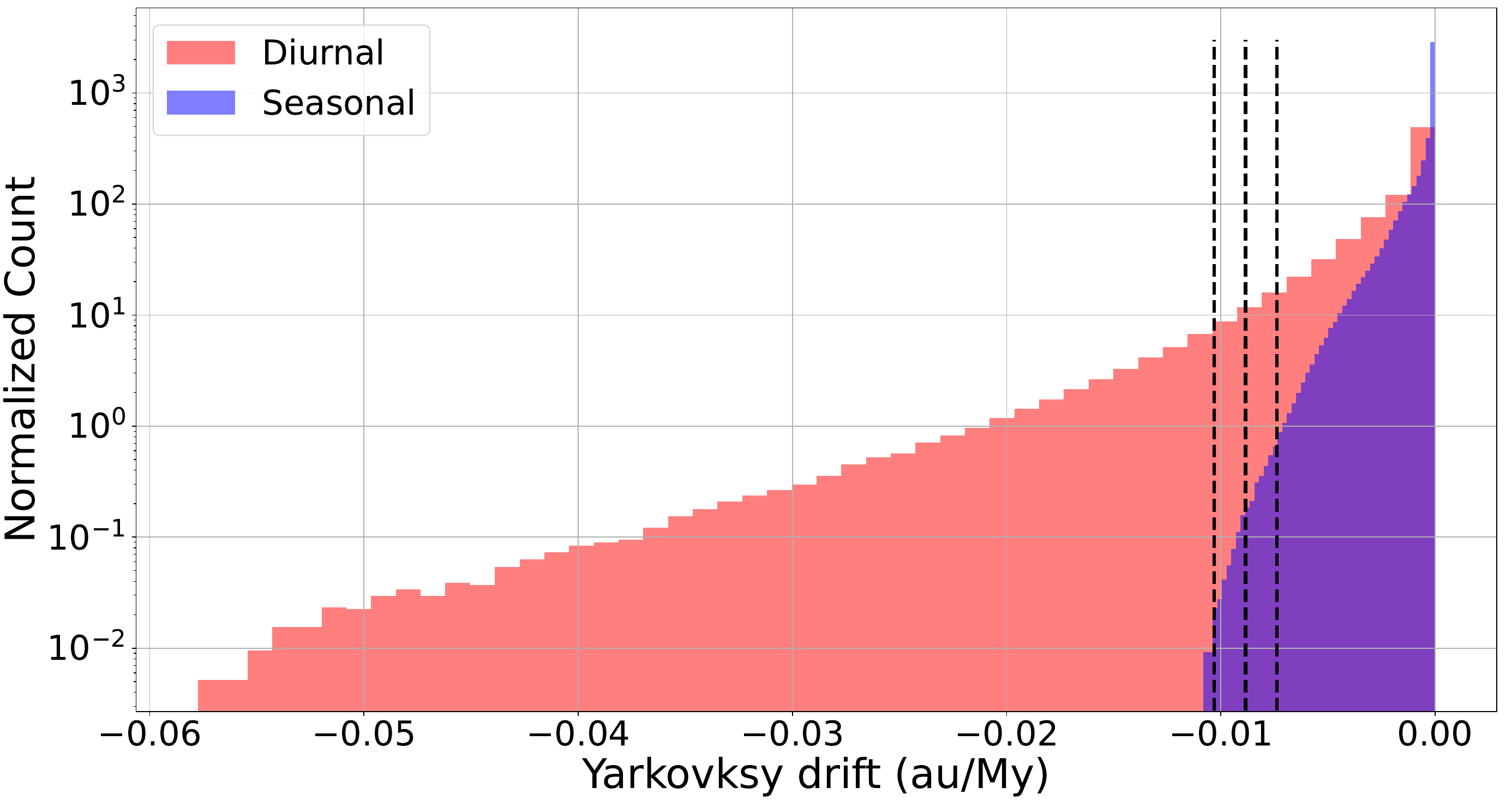}
\caption{Solutions for the diurnal and seasonal components of the Yarkovsky effect for asteroid 2011 PT, shown across plausible ranges of physical parameters. Vertical dashed lines indicate the nominal value and the ±1$\sigma$ interval.}\label{fig:distribution PT}
\end{center}
\end{figure}

\begin{figure}
\begin{center}
\includegraphics[width=\linewidth]{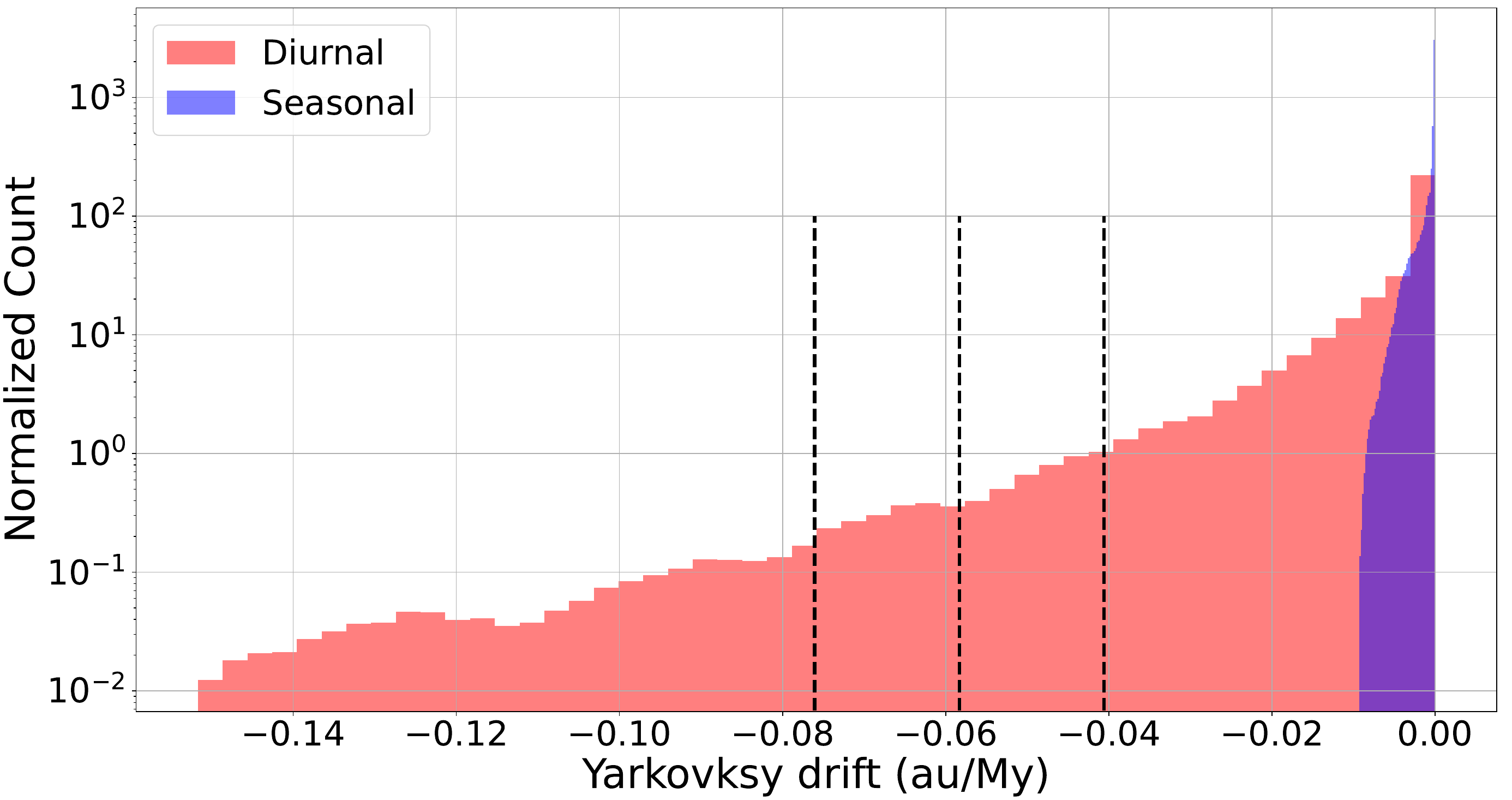}
\caption{Solutions for the diurnal and seasonal components of the Yarkovsky effect for asteroid 2016 GE1, shown across plausible ranges of physical parameters. Vertical dashed lines indicate the nominal value and the ±1$\sigma$ interval.}\label{fig:distribution GE1}
\end{center}
\end{figure}

As shown in Figures \ref{fig:distribution PT} and \ref{fig:distribution GE1}, for asteroid 2011 PT the observed drift can be explained independently by both the diurnal and the seasonal components of the Yarkovsky effect. On the other hand, in the case of 2016 GE1, none of the solutions for the seasonal component fall within the 1-sigma interval. In fact, most are an order of magnitude smaller. Therefore, in the case of this extremely fast rotator, the drift can only be explained by the diurnal component. It should be noted that the distributions shown in these figures represent the solution space over the entire range of physical parameters considered. This does not correspond to a realistic probability distribution, as physical parameters such as size or thermal inertia are not uniformly distributed among asteroids and some values are more probable than others. The purpose of this representation is to demonstrate that, regardless of the probability distribution of these parameters, no combination within the full parameter space yields a sufficient seasonal component drift for 2016 GE1, while both components are possible for 2011 PT. However, the solution space is consistent with the probability density of thermal parameters for both 2011 PT \citep{2021A&A...647A..61F}, where 99\% of the solutions fall below a thermal inertia of 300~J\,m$^{-2}$\,K$^{-1}$\,s$^{-1/2}$, and 2016 GE1 \citep{2023A&A...675A.134F}, where 99\% of the solutions fall below 100~J\,m$^{-2}$\,K$^{-1}$\,s$^{-1/2}$.

\begin{figure}
\begin{center}
\includegraphics[width=\linewidth]{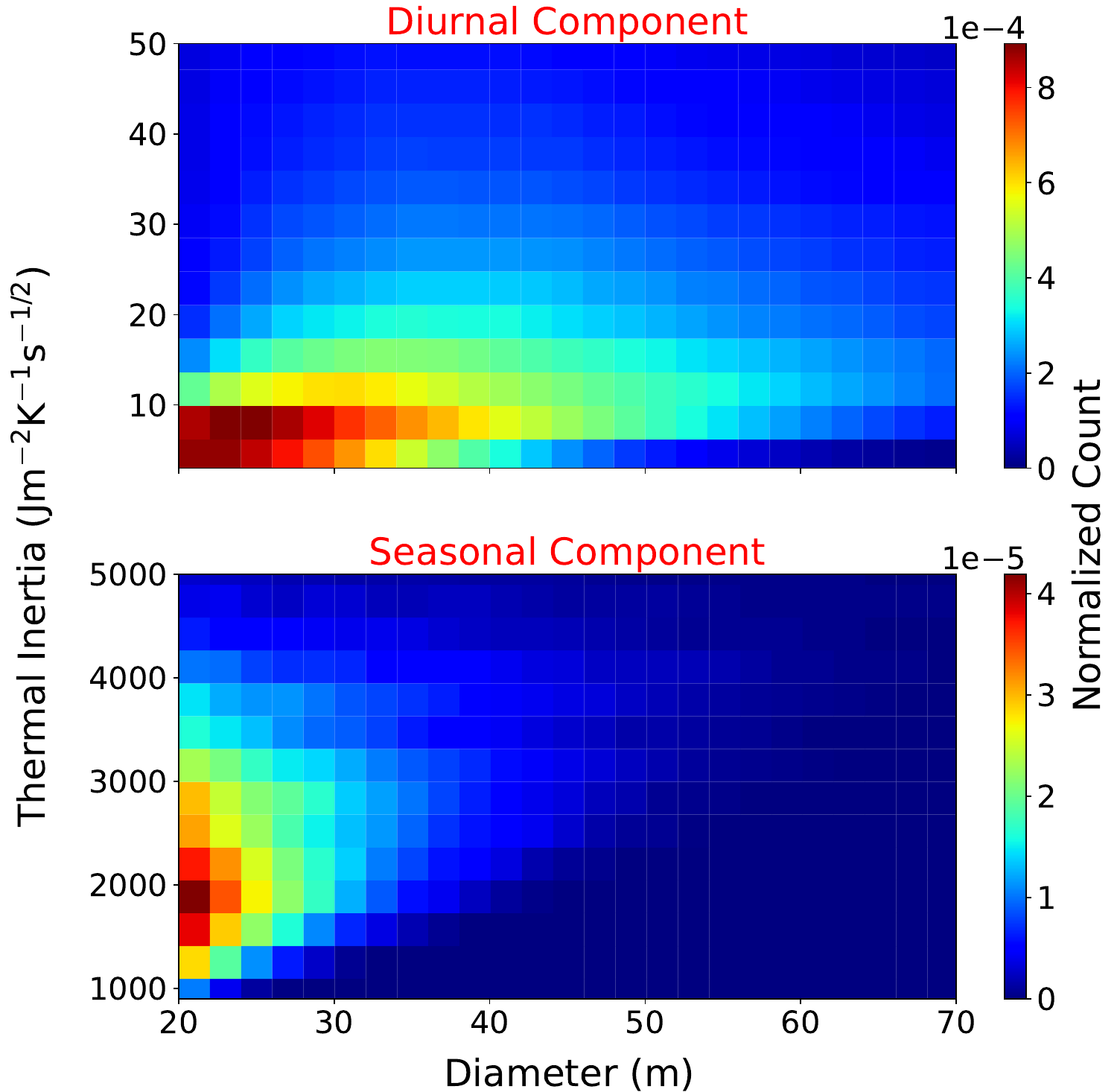}
\caption{Distribution of physical parameters for asteroid 2011 PT resulting in diurnal (upper panel) and seasonal (lower panel) components of the Yarkovsky effect falling within $\pm 1\sigma$ from the nominal drift.
}\label{fig:distribution PT seasonal diurnal}
\end{center}
\end{figure}

As shown in Fig.~\ref{fig:distribution PT seasonal diurnal}, both the seasonal and diurnal components of the Yarkovsky effect can be realized predominantly for smaller sizes, but under very different thermal surface properties. While the diurnal component is primarily associated with low thermal inertia, mostly below 50~J\,m$^{-2}$\,K$^{-1}$\,s$^{-1/2}$, the seasonal component, on the other hand, is typically achieved with much higher thermal inertia, with 99\% of the solutions exceeding 1000~J\,m$^{-2}$\,K$^{-1}$\,s$^{-1/2}$.

\begin{figure}
\begin{center}
\includegraphics[width=\linewidth]{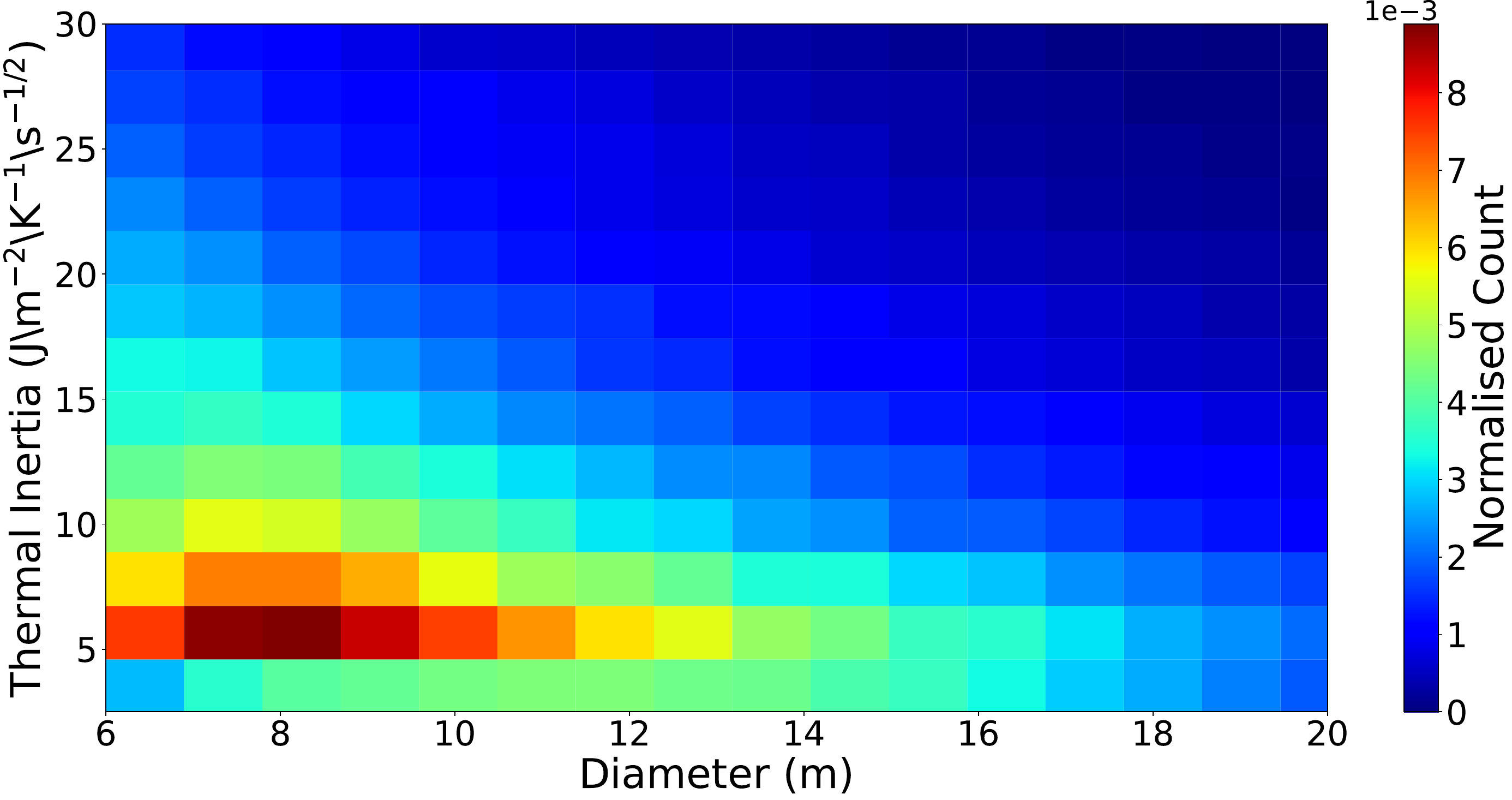}
\caption{Distribution of physical parameters for asteroid 2016 GE1 resulting in diurnal component of the Yarkovsky effect falling within $\pm 1\sigma$ from the nominal drift for the nominal rotation period of 34 seconds.}\label{fig:distribution GE1 parametri}
\end{center}
\end{figure}

\begin{figure}
\begin{center}
\includegraphics[width=\linewidth]{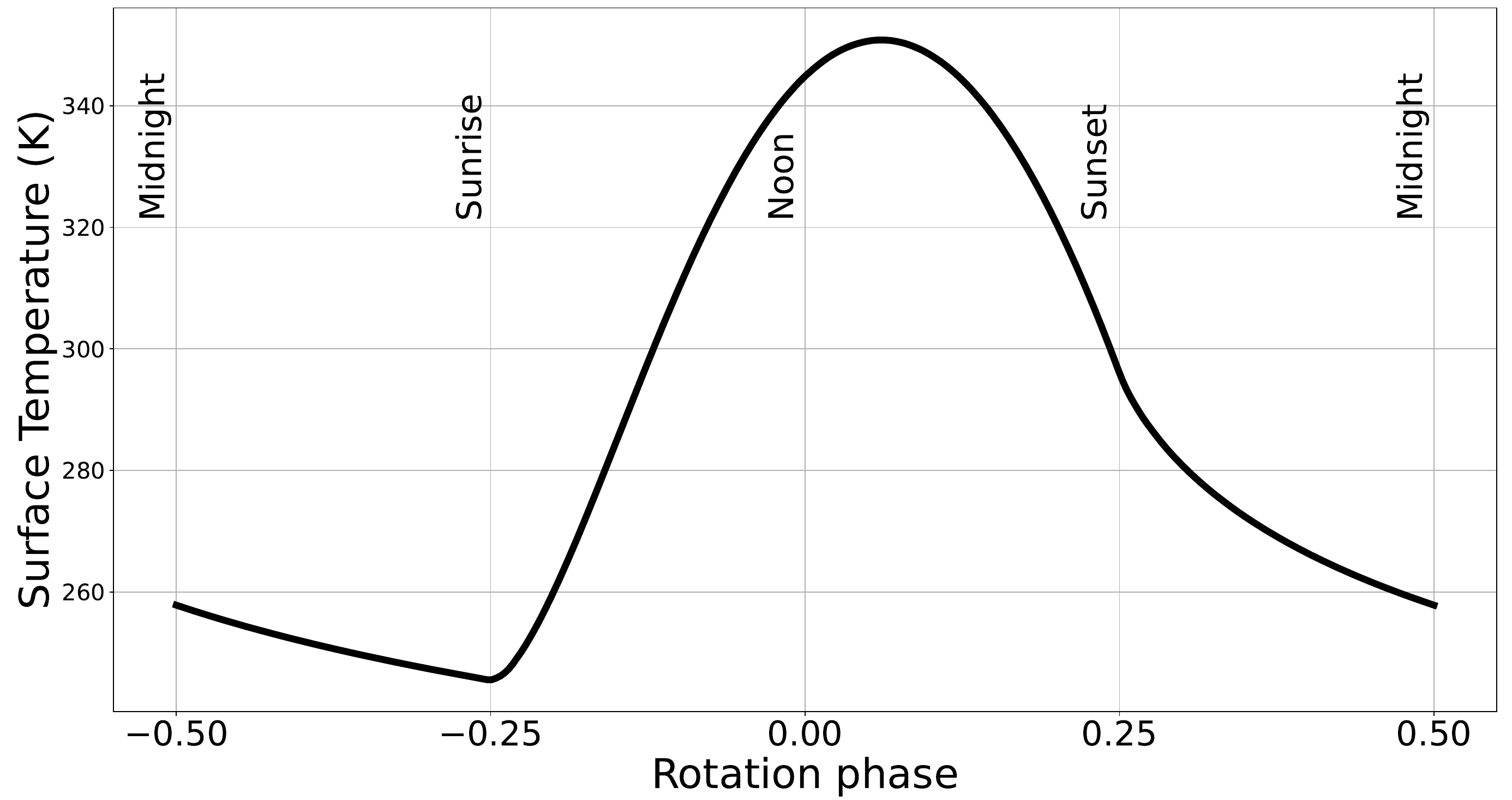}
\caption{Diurnal temperature cycle for the most probable parameters of asteroid 2016~GE1.}\label{fig:GE1 diurnal cycle}
\end{center}
\end{figure}

As shown in Fig.~\ref{fig:distribution GE1 parametri}, the measured Yarkovsky drift for asteroid 2016~GE1 can be reproduced only with extremely low thermal inertia, with 99\% of the cases falling below 100~J\,m$^{-2}$\,K$^{-1}$\,s$^{-1/2}$. To analyze the temperature field required to produce such a drift, the diurnal temperature variations on the surface and in the shallow subsurface of asteroid 2016 GE1 were calculated for the most probable set of physical parameters \citep{2023A&A...675A.134F}: diameter $D = 12~\mathrm{m}$, thermal inertia $TI = 20~\mathrm{J,m^{-2} K^{-1} s^{-1/2}}$, and rotation period $P = 34~\mathrm{s}$. As shown in Figure \ref{fig:GE1 diurnal cycle}, generating the observed Yarkovsky drift requires a very large temperature variation during rotation, reaching up to 100~K at the equator. 

\begin{figure}
\begin{center}
\includegraphics[width=\linewidth]{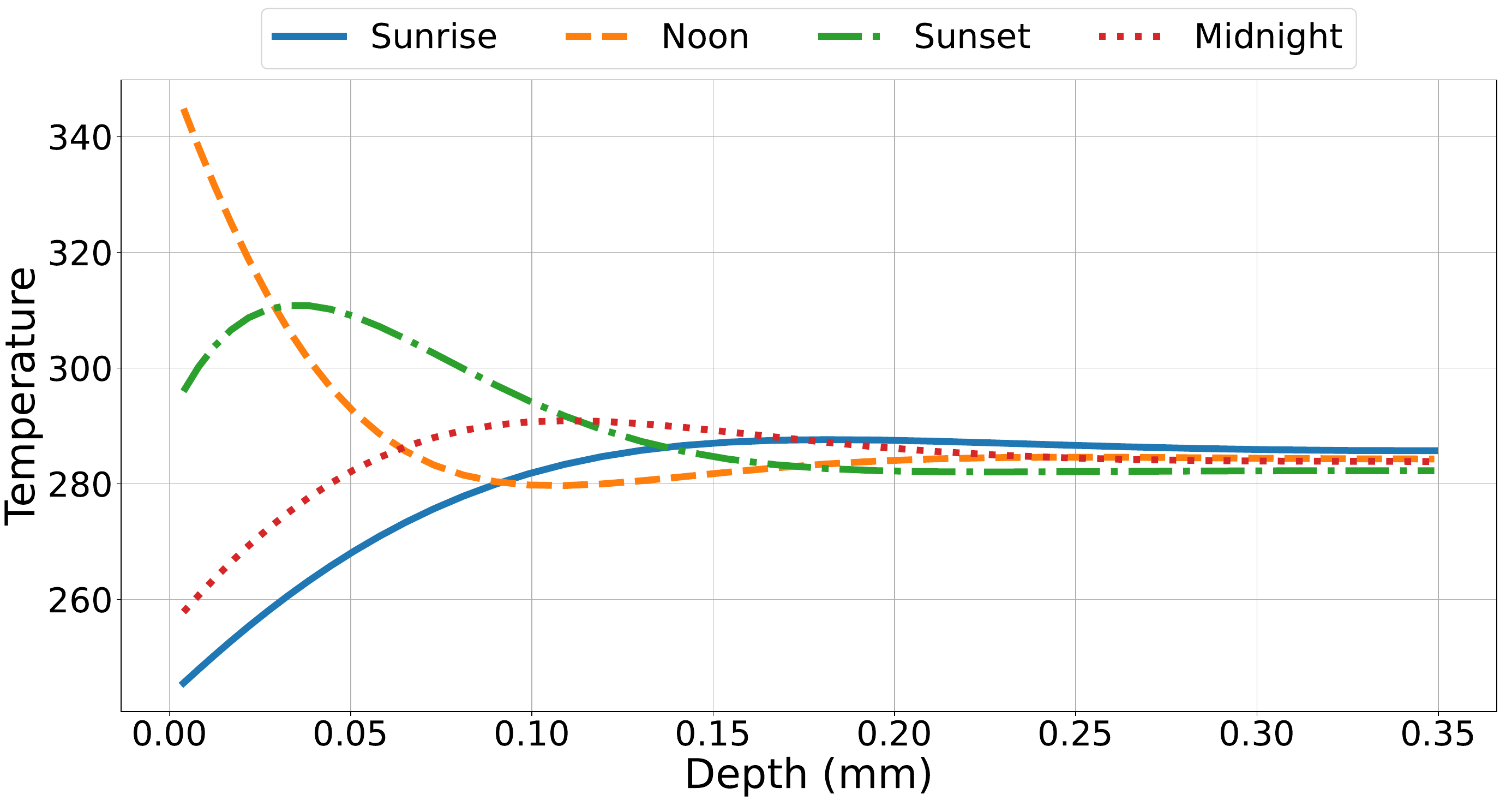}
\caption{Subsurface temperature profiles of asteroid 2016 GE1 for the most probable parameters at four characteristic rotational phases.}\label{fig:GE1 depth profile}
\end{center}
\end{figure}

To further illustrate the thermal response of the surface and subsurface, Figure~\ref{fig:GE1 depth profile} shows the temperature profiles with depth for the most probable parameters of GE1 at four characteristic rotational phases. The results reveal that the extreme temperature cycle required to produce the observed Yarkovsky drift is confined to a very shallow subsurface layer. Specifically, Figure~\ref{fig:GE1 depth profile} demonstrates that this temperature variation occurs with a thin insulating layer on the order of 0.1~mm.

\section{Summary and Conclusions}
\label{sec:summary}

We have developed, implemented, and tested a new 3-D numerical framework that couples high-resolution heat conduction calculations with full-vector radiation recoil force, enabling reliable Yarkovsky effect predictions across all rotation regimes, including sub-minute super-fast rotators.  

The principal outcomes of this work are:
\begin{enumerate}
  \item The Python implementation of the numerical Yarkovsky effect, released as open-source, offers separate modes for the diurnal and seasonal Yarkovsky components.

  \item Benchmark validation for a 100~m spherical test asteroid, the numerical model reproduces THERMOBS surface temperatures to within $\sim$3~K and tracks the analytical diurnal Yarkovsky solution of \citet{vokrouhlicky_1998} to better than $~10\%$. The seasonal component shows agreement better than $~20\%$ over five decades in thermal conductivity.

  \item Treating the numerical solution as ground truth, we probed the robustness of the analytical formula at extreme spin rates, including periods as short as 10 s. Relative differences remain within $\pm10\%$ for $k=10^{-5}$ – 1 W m$^{-1}$ K$^{-1}$, confirming that the analytical theory retains practical accuracy even for super-fast rotators.

 \item We used the numerical model in two asteroid case studies: 2011 PT and 2016 GE1. Monte-Carlo sweeps across plausible physical parameters show that the observed drift of 2011 PT can be matched by either the diurnal or the seasonal component. In contrast, for the 34s rotator 2016 GE1, only the diurnal term reaches the measured value. Explaining the latter requires a thermal inertia $\Gamma\lesssim100$ J m$^{-2}$ K$^{-1}$ s$^{-1/2}$, consistent with a $\sim$0.1 mm insulating layer.

\end{enumerate}

The analysis presented here assumes a spherical shape. For super-fast rotators such as 2011 PT and 2016 GE1 this is a reasonable first-order approximation. Nevertheless, \citet{capek2007} showed that realistic irregular shapes can reduce the diurnal Yarkovsky drift by up to $15\%$ relative to an equal-mass sphere. Applying that correction would affect our drift curves a little, but all conclusions would remain intact.

Surface roughness is likewise not yet included in the present model. Numerical experiments by \citet{2012MNRAS.423..367R} indicate that roughness typically enhances the Yarkovsky effect by $\approx20\%$ and, under very low thermal inertia, by up to a factor of 2. Incorporating roughness would therefore increase, rather than diminish, the drift rates reported here, further supporting our conclusion that super-fast rotators can experience substantial non-gravitational acceleration.

Finally, for asteroid 2016 GE1, the combination of low thermal inertia and a 34-s spin produces temperature excursions of order 100 K on time scales of only tens of seconds. The mechanical response of regolith to such rapid thermal cycling remains unexplored; the thermal fatigue-induced fracturing of exposed rock \citep{2014Natur.508..233D,2024NatCo..15.6206L} could play a crucial role. This is also highly relevant for the mechanical properties of regolith and the dynamics of asteroids' surfaces \citep[e.g.][]{2024A&A...684A.172D}. Quantifying this feedback between surface breakdown and radiative forces is an engaging direction for future work.

\begin{acknowledgements}
      This research was supported by \emph{ The Science Fund of the Republic of Serbia} through Project No. 7453 \emph{Demystifying enigmatic visitors of the near-Earth region (ENIGMA)}. 
\end{acknowledgements}

\bibliographystyle{aa}
\bibliography{enigma}

\begin{thebibliography}{47}
\expandafter\ifx\csname natexlab\endcsname\relax\def\natexlab#1{#1}\fi

\bibitem[{{Bottke} {et~al.}(2006){Bottke}, {Vokrouhlick{\'y}}, {Rubincam}, \& {Nesvorn{\'y}}}]{bottke-etal_2006}
{Bottke}, William~F., J., {Vokrouhlick{\'y}}, D., {Rubincam}, D.~P., \& {Nesvorn{\'y}}, D. 2006, Annual Review of Earth and Planetary Sciences, 34, 157

\bibitem[{{Cheng} {et~al.}(2018){Cheng}, {Rivkin}, {Michel}, {Atchison}, {Barnouin}, {Benner}, {Chabot}, {Ernst}, {Fahnestock}, {Kueppers}, {Pravec}, {Rainey}, {Richardson}, {Stickle}, \& {Thomas}}]{2018P&SS..157..104C}
{Cheng}, A.~F., {Rivkin}, A.~S., {Michel}, P., {et~al.} 2018, \planss, 157, 104

\bibitem[{{Chesley} {et~al.}(2003){Chesley}, {Ostro}, {Vokrouhlick{\'y}}, {{\v{C}}apek}, {Giorgini}, {Nolan}, {Margot}, {Hine}, {Benner}, \& {Chamberlin}}]{chesley-etal_2003}
{Chesley}, S.~R., {Ostro}, S.~J., {Vokrouhlick{\'y}}, D., {et~al.} 2003, Science, 302, 1739

\bibitem[{{Dai} {et~al.}(2024){Dai}, {Yu}, {Cheng}, {Baoyin}, \& {Li}}]{2024A&A...684A.172D}
{Dai}, W.-Y., {Yu}, Y., {Cheng}, B., {Baoyin}, H., \& {Li}, J.-F. 2024, \aap, 684, A172

\bibitem[{{Danby}(1992)}]{Danby1992}
{Danby}, J. 1992, Fundamentals of Celestial Mechanics (Willman-Bell, Inc.Richmond, Virginia, USA)

\bibitem[{{Davidsson} \& {Rickman}(2014)}]{2014Icar..243...58D}
{Davidsson}, B. J.~R. \& {Rickman}, H. 2014, \icarus, 243, 58

\bibitem[{{Delbo}(2004)}]{2004PhDT.......371D}
{Delbo}, M. 2004, PhD thesis, Free University of Berlin, Germany

\bibitem[{{Delbo} {et~al.}(2014){Delbo}, {Libourel}, {Wilkerson}, {Murdoch}, {Michel}, {Ramesh}, {Ganino}, {Verati}, \& {Marchi}}]{2014Natur.508..233D}
{Delbo}, M., {Libourel}, G., {Wilkerson}, J., {et~al.} 2014, \nat, 508, 233

\bibitem[{{Delbo} {et~al.}(2015){Delbo}, {Mueller}, {Emery}, {Rozitis}, \& {Capria}}]{delbo-etal_2015}
{Delbo}, M., {Mueller}, M., {Emery}, J.~P., {Rozitis}, B., \& {Capria}, M.~T. 2015, {Asteroid Thermophysical Modeling} (University of Arizona Press, Tucson), 107--128

\bibitem[{{Farnocchia} {et~al.}(2021){Farnocchia}, {Chesley}, {Takahashi}, {Rozitis}, {Vokrouhlick{\'y}}, {Rush}, {Mastrodemos}, {Kennedy}, {Park}, {Bellerose}, {Lubey}, {Velez}, {Davis}, {Emery}, {Leonard}, {Geeraert}, {Antreasian}, \& {Lauretta}}]{2021Icar..36914594F}
{Farnocchia}, D., {Chesley}, S.~R., {Takahashi}, Y., {et~al.} 2021, \icarus, 369, 114594

\bibitem[{{Fenucci} {et~al.}(2023){Fenucci}, {Novakovi{\'c}}, \& {Mar{\v{c}}eta}}]{2023A&A...675A.134F}
{Fenucci}, M., {Novakovi{\'c}}, B., \& {Mar{\v{c}}eta}, D. 2023, \aap, 675, A134

\bibitem[{{Fenucci} {et~al.}(2021){Fenucci}, {Novakovi{\'c}}, {Vokrouhlick{\'y}}, \& {Weryk}}]{2021A&A...647A..61F}
{Fenucci}, M., {Novakovi{\'c}}, B., {Vokrouhlick{\'y}}, D., \& {Weryk}, R.~J. 2021, \aap, 647, A61

\bibitem[{{Harris}(1998)}]{1998Icar..131..291H}
{Harris}, A.~W. 1998, \icarus, 131, 291

\bibitem[{{Jackson} \& {Rozitis}(2024)}]{2024MNRAS.534.1827J}
{Jackson}, S.~L. \& {Rozitis}, B. 2024, \mnras, 534, 1827

\bibitem[{{Krause} {et~al.}(2011){Krause}, {Blum}, {Skorov}, \& {Trieloff}}]{2011Icar..214..286K}
{Krause}, M., {Blum}, J., {Skorov}, Y.~V., \& {Trieloff}, M. 2011, \icarus, 214, 286

\bibitem[{{Lagerros}(1996)}]{1996A&A...310.1011L}
{Lagerros}, J.~S.~V. 1996, \aap, 310, 1011

\bibitem[{{Lebofsky} \& {Spencer}(1989)}]{1989aste.conf..128L}
{Lebofsky}, L.~A. \& {Spencer}, J.~R. 1989, in Asteroids II, ed. R.~P. {Binzel}, T.~{Gehrels}, \& M.~S. {Matthews}, 128--147

\bibitem[{{Lucchetti} {et~al.}(2024){Lucchetti}, {Cambioni}, {Nakano}, {Barnouin}, {Pajola}, {Penasa}, {Tusberti}, {Ramesh}, {Dotto}, {Ernst}, {Daly}, {Mazzotta Epifani}, {Hirabayashi}, {Parro}, {Poggiali}, {Campo Bagatin}, {Ballouz}, {Chabot}, {Michel}, {Murdoch}, {Vincent}, {Karatekin}, {Rivkin}, {Sunshine}, {Kohout}, {Deshapriya}, {Hasselmann}, {Ieva}, {Beccarelli}, {Ivanovski}, {Rossi}, {Ferrari}, {Rossi}, {Raducan}, {Steckloff}, {Schwartz}, {Brucato}, {Dall'Ora}, {Zinzi}, {Cheng}, {Amoroso}, {Bertini}, {Capannolo}, {Caporali}, {Ceresoli}, {Cremonese}, {Della Corte}, {Gai}, {Gomez Casajus}, {Gramigna}, {Impresario}, {Lasagni Manghi}, {Lavagna}, {Lombardo}, {Modenini}, {Palumbo}, {Perna}, {Pirrotta}, {Tortora}, {Zannoni}, \& {Zanotti}}]{2024NatCo..15.6206L}
{Lucchetti}, A., {Cambioni}, S., {Nakano}, R., {et~al.} 2024, Nature Communications, 15, 6206

\bibitem[{{Lyster} {et~al.}(2024){Lyster}, {Howett}, \& {Penn}}]{2024EPSC...17.1121L}
{Lyster}, D., {Howett}, C., \& {Penn}, J. 2024, in European Planetary Science Congress, EPSC2024--1121

\bibitem[{{Michel} {et~al.}(2022){Michel}, {K{\"u}ppers}, {Bagatin}, {Carry}, {Charnoz}, {de Leon}, {Fitzsimmons}, {Gordo}, {Green}, {H{\'e}rique}, {Juzi}, {Karatekin}, {Kohout}, {Lazzarin}, {Murdoch}, {Okada}, {Palomba}, {Pravec}, {Snodgrass}, {Tortora}, {Tsiganis}, {Ulamec}, {Vincent}, {W{\"u}nnemann}, {Zhang}, {Raducan}, {Dotto}, {Chabot}, {Cheng}, {Rivkin}, {Barnouin}, {Ernst}, {Stickle}, {Richardson}, {Thomas}, {Arakawa}, {Miyamoto}, {Nakamura}, {Sugita}, {Yoshikawa}, {Abell}, {Asphaug}, {Ballouz}, {Bottke}, {Lauretta}, {Walsh}, {Martino}, \& {Carnelli}}]{2022PSJ.....3..160M}
{Michel}, P., {K{\"u}ppers}, M., {Bagatin}, A.~C., {et~al.} 2022, Planetary Science Journal, 3, 160

\bibitem[{{M{\"u}ller}(2007)}]{2007PhDT.......401M}
{M{\"u}ller}, M.~M. 2007, PhD thesis, Free University of Berlin, Germany

\bibitem[{{Myhrvold}(2018)}]{2018Icar..303...91M}
{Myhrvold}, N. 2018, \icarus, 303, 91

\bibitem[{{Nakano} \& {Hirabayashi}(2023)}]{2023Icar..40415647N}
{Nakano}, R. \& {Hirabayashi}, M. 2023, \icarus, 404, 115647

\bibitem[{{Novakovi{\'c}} {et~al.}(2024){Novakovi{\'c}}, {Fenucci}, {Mar{\v{c}}eta}, \& {Pavela}}]{2024PSJ.....5...11N}
{Novakovi{\'c}}, B., {Fenucci}, M., {Mar{\v{c}}eta}, D., \& {Pavela}, D. 2024, Planetary Science Journal, 5, 11

\bibitem[{{Novakovi{\'c}} {et~al.}(2022){Novakovi{\'c}}, {Vokrouhlick{\'y}}, {Spoto}, \& {Nesvorn{\'y}}}]{2022CeMDA.134...34N}
{Novakovi{\'c}}, B., {Vokrouhlick{\'y}}, D., {Spoto}, F., \& {Nesvorn{\'y}}, D. 2022, Celestial Mechanics and Dynamical Astronomy, 134, 34

\bibitem[{{Noyes} {et~al.}(2022){Noyes}, {Consolmagno}, {Macke}, {Britt}, \& {Opeil}}]{2022M&PS...57.1706N}
{Noyes}, C.~S., {Consolmagno}, G.~J., {Macke}, R.~J., {Britt}, D.~T., \& {Opeil}, C.~P. 2022, Meteoritics Planet. Sci., 57, 1706

\bibitem[{{Rozitis} \& {Green}(2011)}]{2011MNRAS.415.2042R}
{Rozitis}, B. \& {Green}, S.~F. 2011, \mnras, 415, 2042

\bibitem[{{Rozitis} \& {Green}(2012)}]{2012MNRAS.423..367R}
{Rozitis}, B. \& {Green}, S.~F. 2012, \mnras, 423, 367

\bibitem[{{Sakatani} {et~al.}(2017){Sakatani}, {Ogawa}, {Iijima}, {Arakawa}, {Honda}, \& {Tanaka}}]{2017AIPA....7a5310S}
{Sakatani}, N., {Ogawa}, K., {Iijima}, Y., {et~al.} 2017, AIP Advances, 7, 015310

\bibitem[{{Sekiya} \& {Shimoda}(2013)}]{2013P&SS...84..112S}
{Sekiya}, M. \& {Shimoda}, A.~A. 2013, \planss, 84, 112

\bibitem[{{Sekiya} \& {Shimoda}(2014)}]{2014P&SS...97...23S}
{Sekiya}, M. \& {Shimoda}, A.~A. 2014, \planss, 97, 23

\bibitem[{{Sorli} {et~al.}(2025){Sorli}, {Hayne}, {Cueva}, {Long}, {McMahon}, \& {Scheeres}}]{2025Icar..43416527S}
{Sorli}, K.~C., {Hayne}, P.~O., {Cueva}, R.~H., {et~al.} 2025, \icarus, 434, 116527

\bibitem[{{Spitale} \& {Greenberg}(2001)}]{2001Icar..149..222S}
{Spitale}, J. \& {Greenberg}, R. 2001, \icarus, 149, 222

\bibitem[{{Spitale} \& {Greenberg}(2002)}]{2002Icar..156..211S}
{Spitale}, J. \& {Greenberg}, R. 2002, \icarus, 156, 211

\bibitem[{\v{C}apek(2007)}]{capek2007}
\v{C}apek, D. 2007, PhD thesis, Charles University, Faculty of Mathematics and Physics Astronomical Institute, Prague

\bibitem[{{Vokrouhlicky}(1998)}]{1998A&A...335.1093V}
{Vokrouhlicky}, D. 1998, \aap, 335, 1093

\bibitem[{{Vokrouhlick{\'y}}(1998)}]{vokrouhlicky_1998}
{Vokrouhlick{\'y}}, D. 1998, \aap, 338, 353

\bibitem[{{Vokrouhlick{\'y}}(1999)}]{1999A&A...344..362V}
{Vokrouhlick{\'y}}, D. 1999, \aap, 344, 362

\bibitem[{{Vokrouhlick{\'y}}(2006)}]{2006A&A...459..275V}
{Vokrouhlick{\'y}}, D. 2006, \aap, 459, 275

\bibitem[{{Vokrouhlick{\'y}} \& {Bro{\v{z}} }(1999)}]{vokrouhlicky-broz_1999}
{Vokrouhlick{\'y}}, D. \& {Bro{\v{z}} }, M. 1999, \aap, 350, 1079

\bibitem[{Vokrouhlický {et~al.}(2015)Vokrouhlický, Bottke, Chesley, Scheeres, \& Statler}]{vok2015}
Vokrouhlický, D., Bottke, W.~F., Chesley, S.~R., Scheeres, D.~J., \& Statler, T.~S. 2015, The Yarkovsky and YORP Effects (University of Arizona Press), 509--532

\bibitem[{{Warner} {et~al.}(2009){Warner}, {Harris}, \& {Pravec}}]{2009Icar..202..134W}
{Warner}, B.~D., {Harris}, A.~W., \& {Pravec}, P. 2009, \icarus, 202, 134

\bibitem[{{Xu} {et~al.}(2022){Xu}, {Zhou}, {Hui}, \& {Li}}]{2022A&A...666A..65X}
{Xu}, Y.-B., {Zhou}, L.-Y., {Hui}, H., \& {Li}, J.-Y. 2022, \aap, 666, A65

\bibitem[{{Zhang} {et~al.}(2025){Zhang}, {Xu}, {Qi}, {Zhou}, \& {Li}}]{2025arXiv250509482Z}
{Zhang}, Y., {Xu}, Y.-B., {Qi}, Z., {Zhou}, L.-Y., \& {Li}, J.-Y. 2025, arXiv e-prints, arXiv:2505.09482

\bibitem[{{Zhao} {et~al.}(2024){Zhao}, {Lei}, \& {Shi}}]{2024A&A...691A.224Z}
{Zhao}, S., {Lei}, H., \& {Shi}, X. 2024, \aap, 691, A224

\bibitem[{{Zhou}(2024)}]{2024A&A...692L...2Z}
{Zhou}, W.-H. 2024, \aap, 692, L2

\bibitem[{{Zhou} {et~al.}(2024){Zhou}, {Vokrouhlick{\'y}}, {Kanamaru}, {Agrusa}, {Pravec}, {Delbo}, \& {Michel}}]{2024ApJ...968L...3Z}
{Zhou}, W.-H., {Vokrouhlick{\'y}}, D., {Kanamaru}, M., {et~al.} 2024, \apjl, 968, L3

\end{thebibliography}
\end{document}